\documentclass[aps, pre, twocolumn, a4paper, superscriptaddress, floatfix,longbibliography]{revtex4-1}
\usepackage{graphicx}
\usepackage{amsmath,mathtools}
\usepackage{placeins}
\usepackage{color}
\usepackage{hyperref}
\usepackage{url}

\newcommand{\SARS}{SARS-CoV-2 }
\begin{document}

\title{Bi-stability of SUDR+K model of epidemics and test kits applied to COVID-19}


\author{Vinko Zlati\'c}
\affiliation{Division of Theoretical Physics, Ru\-{d}er Bo\v{s}kovi\'c Institute, Zagreb, Croatia}

\author{Irena Barja\v{s}i\'c}
\affiliation{Faculty of Science, Zagreb University, Zagreb, Croatia}

\author{Andrea Kadovi\'c}
\affiliation{University Hospital Centre Zagreb, Zagreb, Croatia}


\author{Hrvoje \v{S}tefan\v{c}i\'c}
\affiliation{Catholic University of Croatia, Ilica 242, 10000 Zagreb, Croatia }

\author{Andrea Gabrielli}
\affiliation{Engineering Department, University ``Roma Tre", Via Vito Volterra 62, 00146 - Rome, Italy}


%

\date{\today}

\begin{abstract}
 Motivated by different responses of governments of many countries to COVID-19, here we develop a toy model of the epidemics dependence on the availability of tests. Our model, that we call SUDR+K, is based on the well known SIR model, but it splits the total fraction of infected individuals into two components: those that are undetected and those that are detected through tests. Moreover, we assume that available tests increase at a constant rate from the beginning of epidemics but are consumed to detect infected individuals. Strikingly we find a bi-stable behavior between a phase with a giant fraction of infected and a phase with a very small fraction. We show that the separation between these two regimes is governed by a match between the rate of testing and the rate of infection spread at given time. We also show that the existence of two phases is largely independent of the precise mathematical form of the term describing the rate at which undetected individuals are tested and detected. The present research implies that a vigorous early testing activity, before the epidemics enters into its giant phase, can potentially keep epidemics under control, and that even a very small change of the testing rate can increase or decrease the size of the whole epidemics over many orders of magnitude. 
\end{abstract}

\maketitle 

\section{Introduction}
The recent outbreak of the \SARS  virus and the associated illness COVID-19 has triggered, in this century, unprecedented containment measures around the world including the complete lock-down of the populations of all towns in in many European and non-European countries, including China and US \cite{world2020report}. The World Health Organization has declared the diffusion of COVID-19 to be a pandemics and issued a strong warning of a severe global threat \cite{WHOWarning}.
In the case of the COVID-19 epidemics there is also an {\em infodemic} of true and false news about the danger, the diffusion and the treatments of COVID-19 \cite{cinelli2020covid19}. This context muddles the attempts to understand the epidemics and confuses people. At the same time we witness a very active participation among scientists following the epidemics on all social media and platforms. Some of important questions are: \emph{(i)} How many infected people are undetected? \emph{(ii)} How the number of tests and testing policies affects the dynamics of epidemics? \emph{(iii)} Is there any evident benefit in early massive testing?\\ Some of those questions have been addressed with different methods in the context of previous different epidemics and have recently been addressed in the present COVID-19 diffusion with almost no explicit effort of theoretical modeling until now. In \cite{nouvellet2015role}, authors statistically evaluate different strategies of testing in the context of ebola epidemics and show the importance of early testing. They found that an appropriate implementation of this practice would reduce epidemics by one third. In \cite{petric2006role} authors review laboratory testing for influenza, which is often mentioned to be similar to \SARS in spreading features, and lay out all the possible ways in which early tests can be used to fight the diffusion of such a disease. In \cite{Lieabb3221} authors conclude that undocumented infections present the same main channels of geographic spread of \SARS.

There is an ongoing effort in calibrating models for the dynamics of this epidemics in order to set the values of the model parameters significantly affecting the diffusion \cite{CDCinfo,sanche2020novel,zhao2020preliminary}. In this letter we  adopt the available numerical estimates published in this studies. Parameters, whose calibration is impossible due to lack of data, are implicitly kept within realistic ranges. 

In order to explicitly take into account the different impacts on the spreading dynamics of undetected and detected infected individuals, and the contribution of the available number of testing kits to put the epidemics under control, here we extend the usual SIR model to a novel ``SUDR + K" one \footnote{Here we adopt SIR model instead of SIS (susceptible-infected-susceptible) as we assume immunization for recovered patients even though this is not yet clinically proved. However we can ground on the clinical observation that other coronaviruses affecting humans lead at least to a partial immunization of recovered individuals {\tt https://www.nature.com/articles/d41586-020-00798-8}}. In the model we propose four states for the individuals of a population: $S$ (susceptible), $U$ (undetected), $D$ (detected) and $R$ (removed). Moreover, we introduce an additional variable $K$ which models the number of available test kits. Susceptible are those individuals in the population who can acquire disease. Infected individuals can be detected or undetected, therefore $I=U+D$. Detected are those that are positively tested, and undetected are infected of which no one precisely knows of (although some may be suspected for infection). Removed are those individuals that either healed and acquired immunity or are deceased. Total number of people in the population is $N$.
Lower case letters represent fraction of population, $s+u+d+r=1$ ($u+d=i=I/N$), and $k=K/N$ represents available number of tests per capita. 

Even though in reality there are different kinds of tests (including nasopharyngeal and oropharyngeal swabs, bronchoalveolar lavage, serum testing, CT scan etc. \cite{Tests}), we gather all the kinds in a single family of tests.


The model we propose is defined by the following equations:


\begin{align}
    \dot{s}&=-\beta s u \label{1} \\
    \dot{u}&=\beta s u -\delta u k - \gamma u \label{2} \\
    \dot{d}&=\delta u k - \gamma d  \label{3} \\ \dot{r}&=\gamma(u+d) \label{4} \\ 
    \dot{k}&=\alpha-\epsilon\delta u k \label{5}
\end{align}

Equation \eqref{1} is just the usual equation of SIR model that represents the dynamics from susceptible to be infected after exposure. Here we put $u$ instead of $i$, because we assume that after detection the probability to spread the contagion becomes negligible \footnote{This assumption is reasonable in a condition of efficient and reliable health care system, but if hospitals are under stress or have not enough protective gear one could actually even expect significant contribution from detected individuals.}.


Equation \eqref{2} needs a more detailed explanation. The first term just represents the fraction of individuals that changed their state from susceptible to infected. The second term models the change of undetected to detected by testing. If there are no tests no one can get detected, if there are no undetected again no one can get detected. It is then proportional to both the numbers of undetected and of kits. It is motivated by the idea that infected individuals report to hospital on the basis of symptoms (proportional to $u$) and get tested with higher probability if there is abundance of kits or lower if there is a scarcity of kits. The third term represents just the fraction of individuals that gets removed without ever been detected. 
Equation \eqref{3} has the first term of opposite sign with respect to the analogous term in the previous equation and an additional removal term of detected individuals. Although the removal of an undetected individual happens only through healing (direct death without a transition to $d$ can be as a first approximation neglected), while the removal of a detected individual can be due to both healing and death, we chose to remove detected and undetected individuals with equal rate, leading to Eq.~\eqref{4}, to reduce the number of parameters.  
Equation \eqref{5} represents the dynamics of the available number of kits. The first term in the equation represents a constant growth of the number of kits (fixed production of kits per day). The second term reasonably assumes that kits are used proportionally to the number of undetected individuals and the number of available kits, and also prevents the number of kits to become negative. 
The parameter $\epsilon>1$ measures how many more tests have to be done to switch an undetected individual to detected, so that the corresponding term in the equation has to be equal or larger than the corresponding term $\delta u k$ in Eqs. \eqref{2} and \eqref{3}.

Of course there can be higher order contributions in all equations, however in our opinion Eqs.~\eqref{1}-\eqref{5} are the simplest possible to get plausible dynamics.

\begin{figure}[t!]
\vspace{-2 cm}
\includegraphics[width=.91\columnwidth]{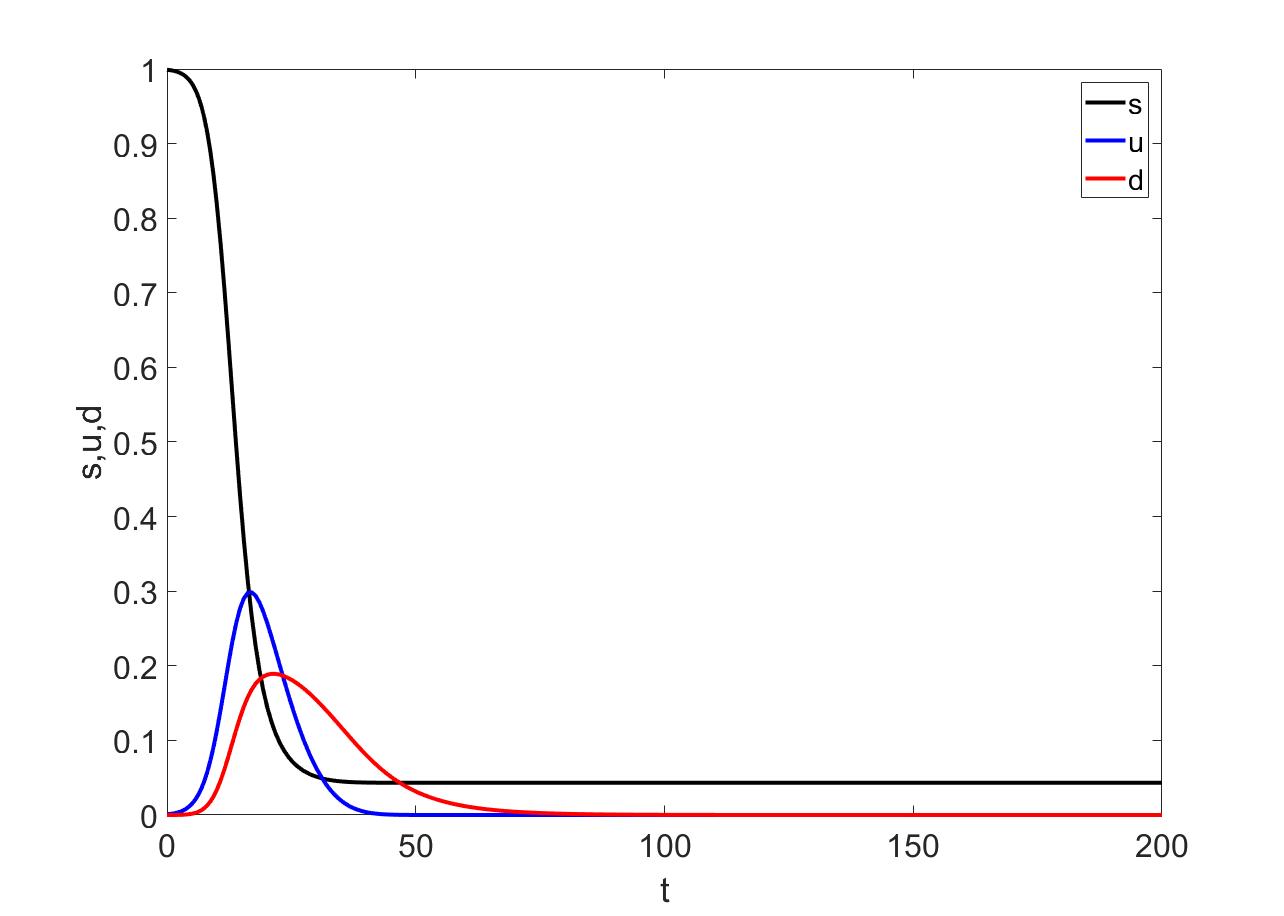}
\centering
\vspace{0 cm}
\caption{Red line represents detected fraction of population, through time; blue - undetected and black - susceptible for parameters $\beta = \ln{2}$, $\delta=0.5$, $\gamma=\ln{2}/7$, $\alpha=0.02$, $\epsilon=1$.}
\label{Fig: Fig1}
\end{figure}

\section{More general models for detection}
\label{models}

An alternative model for detection can be obtained in the following way.
First, let us assume that for each time increment $\Delta K=\alpha$ new kits are produced, and that a fraction $0<\delta'<1$ of available kits $K$ is used for people accepted in the hospitals. 
This means that the number of kits used on hospitalized people is $\delta' K$.
On the other hand the fraction of people arriving at hospitals with symptoms is proportional to fraction of undetected therefore $\delta'=\delta u$. 
Moreover, let us assume that each of these newly detected individuals had previously infected other $\beta s$ susceptible individuals ($\beta <1$). Therefore we could expect that the number of newly detected is 
\begin{align}
    \Delta D &=\delta u K (1+\beta s)=u\Phi(K,s\delta,\beta)\\
    \Delta d& =\Delta D/N = u\phi(k,s,\delta,\beta)\,.
\end{align}
where
\[\phi(k,s,\delta,\beta)=\Phi(K,s\delta,\beta)/N=\delta k(1+\beta s)\,.\]
Consequently, the model equations now become:
\begin{align}
\dot{s} &= -\beta s u \label{1a} \\
    \dot{u} &= \beta s u - u \phi - \gamma u \label{2a} \\
    \dot{d} &= u\phi - \gamma d  \label{3a} \\
    \dot{r} &= \gamma(u+d) \label{4a}\\
    \dot{k} &= \alpha-\epsilon  \phi\,. \label{5a}
\end{align}

Alternative and more complex coupling terms between detected and undetected individuals, leading to different coupling functions $\phi$, can in principle be possible. Let us call $\delta$ to be the parameter that describes the efficiency of the detection process in all different models.
Two possibilities are
\begin{align}
    u\phi(u^{\delta-1},k)=u^{\delta}k\,,
\end{align}

which is a term often used in chemical kinetics in $A+B\rightarrow C$ \cite{gardiner1985handbook}, and
\begin{align}
    u\phi(k,s,u,\delta)=k\frac{u}{\delta s+u}
\end{align}
which is also typical in the kinetics of chemical reactions \cite{gardiner1985handbook}.  As $\delta<1$, the interpretation is that each unit of kits will be used on either susceptible or undetected people, but undetected individuals are more probable to be tested, therefore $\delta$ reduces the susceptible cohort.  The number of  new detected subjects in a single time-step is given by the ratio between $u$ and the total fraction of individuals subjected to tests (which can be either susceptible or undetected). The rate of finding is then proportional to $k$ and this factor. 

In Eq.~\eqref{3a} we have assumed that spreading of disease and testing happen at the same time. More realistic model could include expected incubation time $\tau$ and then the term would become delayed with 
\begin{align}
    \delta u k (1+\beta s(t-\tau))&=u\phi(k,s,\beta,\tau).
\end{align}
All the above terms can be collected in a single function
 \begin{align}
     u\phi(u,k,\delta,\beta,s)\,,
     \label{FI}
 \end{align}
that we will use in the following.

\section{Results}

\begin{figure}[t!]
\includegraphics[width=.91\columnwidth]{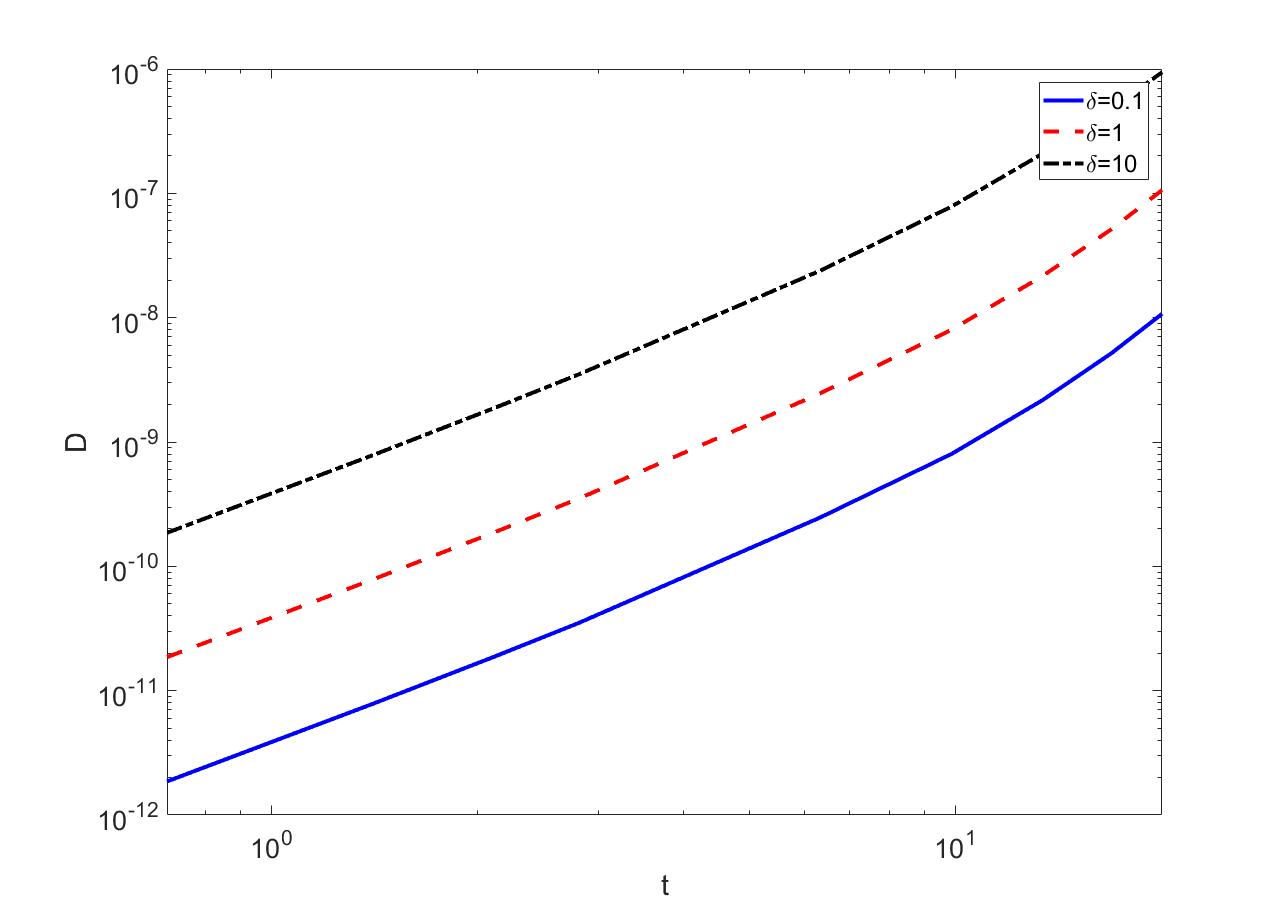}
\centering
\caption{Fraction of detected individuals for different values of parameter $\delta$ , using $\beta=ln(2)/2.7$,$\gamma=ln(2)/7$, $\alpha=7.1429\cdot10^{-5}$, $\epsilon=2$. In the initial stage of the growth we see a power law onset of epidemics. }
\label{Fig: Italyloglog}
\end{figure}

Whichever model we choose, we qualitatively observe the same qualitative behavior. We therefore report in the figures the results for the original model defined by Eqs.~\eqref{1}-\eqref{5}. Generally speaking, we find a difference both in size and temporal position between the two peaks of detected and undetected individuals, as shown in Figure \ref{Fig: Fig1}. Depending on the values of the parameters, sizes and time ordering of the two peaks vary, but the peak of undetected individuals is always higher than the peak of detected ones. 

In Figure \ref{Fig: Italyloglog} we show that, for the chosen values of the parameters, at the very beginning the growth of detected subjects is well fit by a power-law with exponent $\approx 2$ in substantial agreement with results by Maier and Brockman \cite{maier2020effective}. The reason for this initial behavior is very similar to what studied in their model in the sense that there is a reduction of the epidemic spreading for those individual that enter into this new compartment.  This is also checked from the analytical point of view and an expression very similar to the one found in \cite{maier2020effective} is obtained.
Indeed at the start of epidemics, we can safely assume $s\approx 1$ in Eq.~\eqref{2}.

\begin{figure}[h!]
\includegraphics[width=.91\columnwidth]{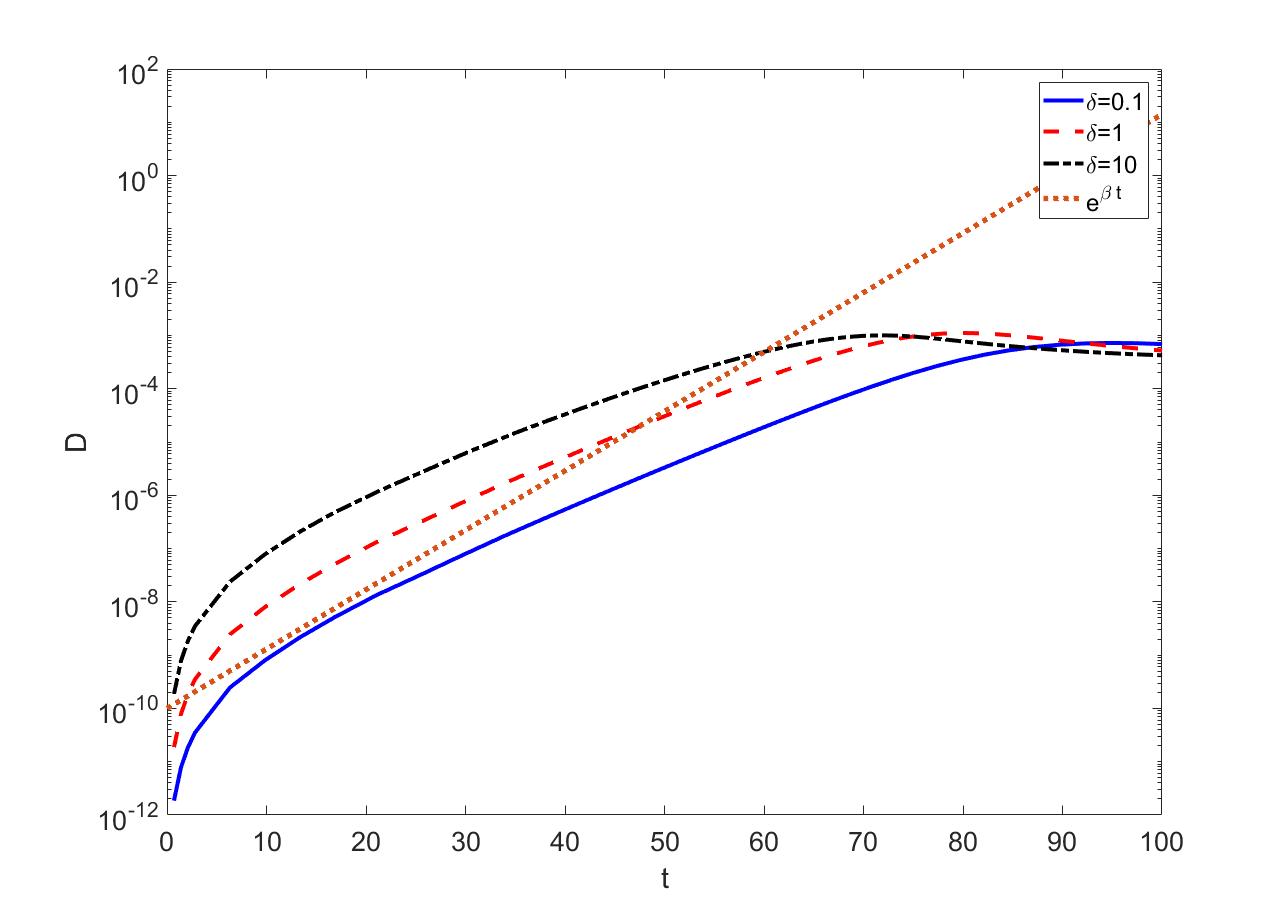}
\centering
\caption{Fraction of detected individuals as function of time for parameters $\beta = \ln{2}/2.7$, $\gamma=\ln{2}/7$, $\alpha=7.1429\cdot10^{-5}$, $\epsilon=2$. After an initial power-law increase, the epidemics shows an exponential increase. Notice that the exponential for the detected is lower than the one which would describe beginning of epidemic in SIR model.}
\label{Fig: Italylinlog}
\end{figure}

However one can see that the fraction of infected individuals in such a power-law regime, multiplied for instance by the Italian population predicts less than one single individual, and therefore this very initial theoretical regime is unobserved in real data for practically all countries. On the contrary, in Figure \ref{Fig: Italylinlog} one can see that for a wide range of parameters values a successive exponential growth is obtained as expected in any epidemics diffusion. In this respect it is noteworthy that SIR model at the start of epidemics exhibits an exponential increase of infected $\sim e^{(\beta-\gamma)t}$, while the number of undetected in our SUDR+K model at this stage of  the epidemics grows as $\sim e^{(\beta-\phi-\gamma)t}$. The fraction of detected individuals on the other hand grows slower than the infected ones in comparable SIR model, thus possibly significantly affecting the measurement of epidemic parameters. . 


\begin{figure}[h!]
\includegraphics[width=.91\columnwidth]{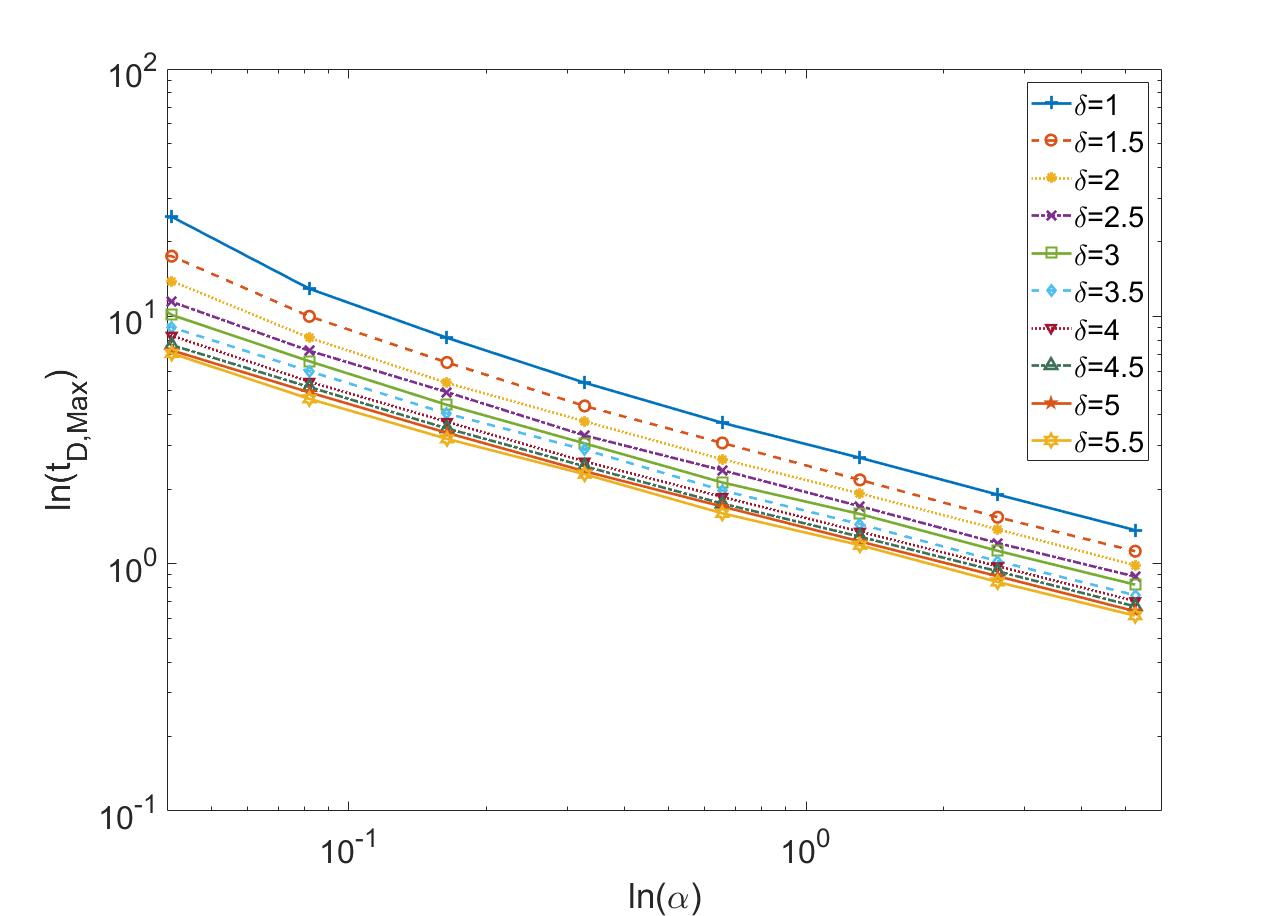}
\centering
\caption{Scaling behavior of the peak time $t_{D,max}$ of detected fraction with growth rate $\alpha$ of testing kit production. The other parameters are set as it follows: $\beta=\ln{2}$ and $\gamma=0.099$, $\epsilon=2$. The scaling can be well fitted by a power law \eqref{tdmax}.
}
\label{Fig: timepeak}
\end{figure}

One of the most interesting aspects of our new model is the appearance of two different peaks in the dynamical evolution of the fractions of the two sub-classes of infected people, undetected and detected.  The peak related to undetected individuals is in general occurring before the peak of detected ones. 
The earlier the peak of detected happens, the smaller the number of total infected at the end of the epidemics. We have found a very interesting relationship between the time $t_{D,max}$ at which the peak of detected occurs and the parameter $\alpha$ giving the production rate of the testing kits:
\begin{align}
    t_{D,max}\sim\alpha^{-\eta}.
    \label{tdmax}
\end{align} 
The value of exponent in Figure \ref{tdmax} is approximately $\eta\approx 2$, but different values of it are found for different values of the other parameters.
The power law relation is very clear in Figure \ref{Fig: timepeak}. In all cases the higher $\alpha$, the smaller $t_{D,max}$.

\begin{figure}[t!]
\includegraphics[width=.91\columnwidth]{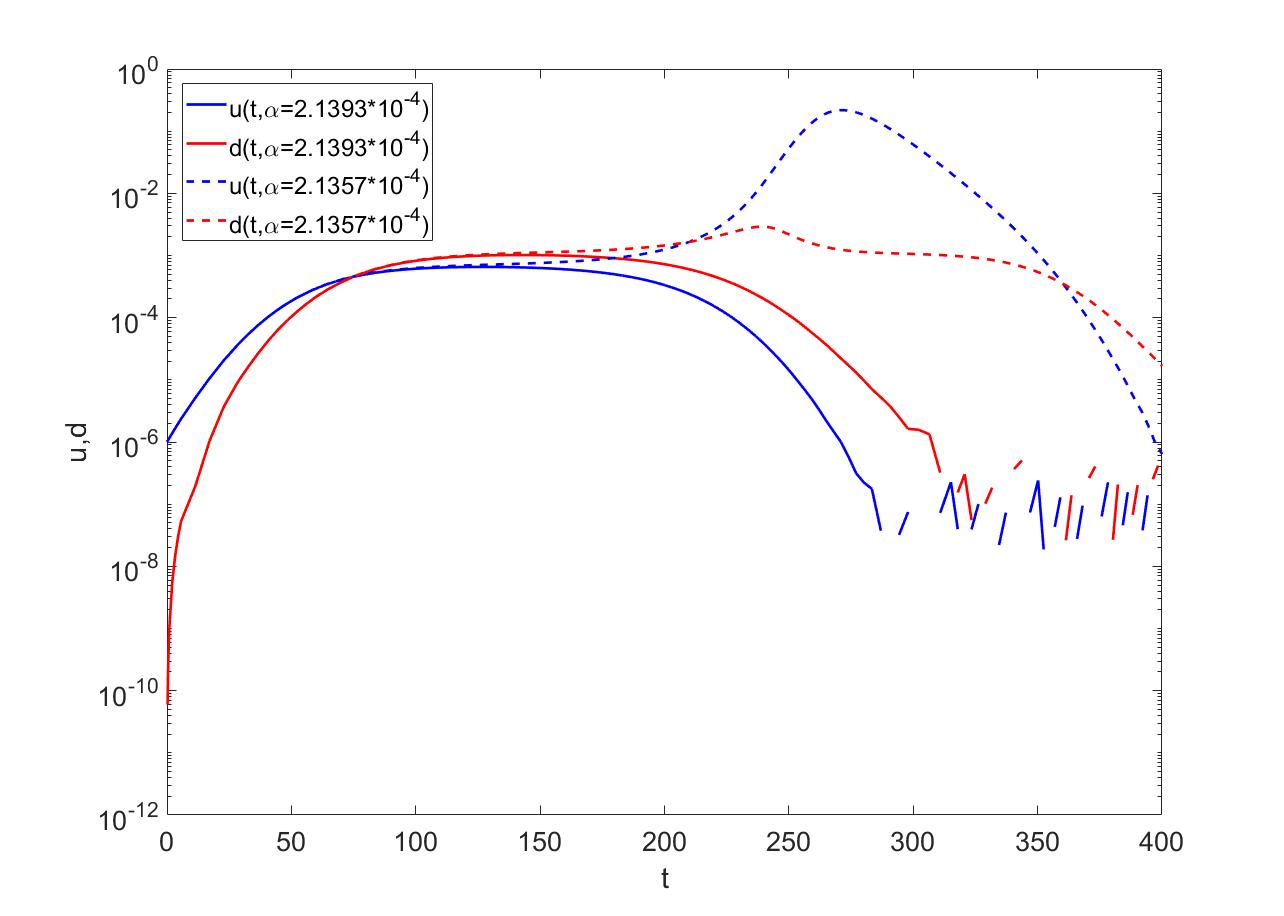}
\centering
\caption{For the usual choice of parameters $\beta=\ln{2}/2.7$, $\gamma=0.099$, $\delta=10$, $\epsilon=2$, we see bi-stability of the time evolution of the epidemic diffusion, fractions $u$ and $d$, at two different but very close values of the rate $\alpha$. We observe a strong response of the system jumping from a phase of full blown epidemics to an almost disappearing one.}
\label{Fig: bistability alpha}
\end{figure}

\begin{figure}[t!]
\includegraphics[width=.91\columnwidth]{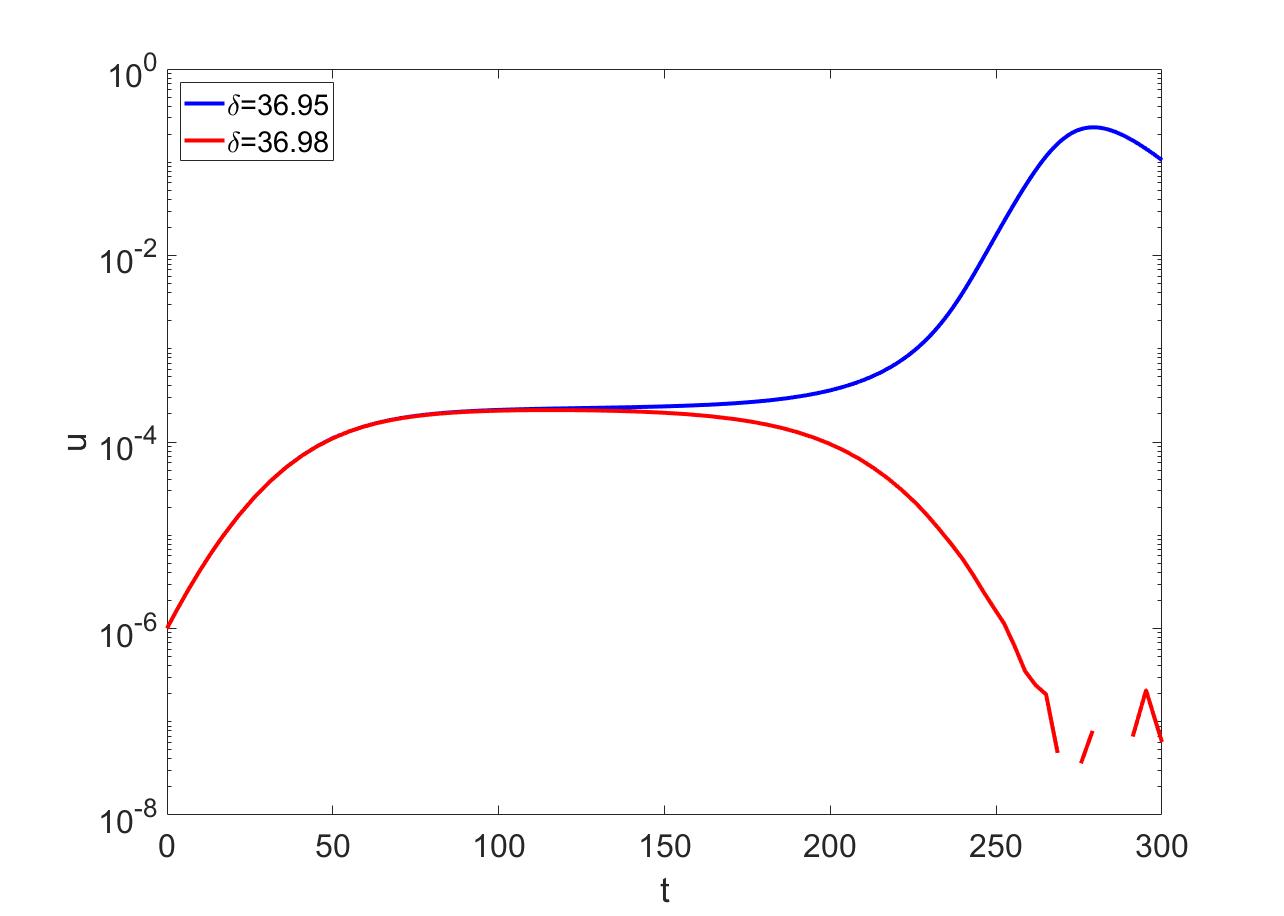}
\centering
\caption{Fraction of undetected infected individuals $u$ as function of time at two very close values for $\delta$, while the other parameter are set at the following values: $\beta=\ln{(2)}/2.7$, $\gamma=0.099$, $\alpha=7.14\cdot10^{-5}$, $\epsilon=2$. We again observe bi-stability and a strong response of the system to small variations of $\delta$ determining an abrupt jump from a phase of full blown epidemics to an almost disappearing one.}
\label{Fig: bistability delta}
\end{figure}

However, the most striking result of our model is represented by Figures \ref{Fig: bistability alpha} and \ref{Fig: bistability delta}.

In Figure \ref{Fig: bistability alpha}, for very realistic values of $\alpha$, we observe a switching behavior between two phases, one with a full blow epidemics, and the other one in which the epidemic diffusion practically disappears before the development of a macroscopic spreading across the population. The separation between this two different behaviors appears to be a real bifurcation point. Indeed, by fixing all other parameters, we observe the switching between such an explosive and a self-contained behaviors for values of $\alpha$ which are very close one to each other. This strongly suggests the possibility of a huge effect on the epidemics diffusion even for a change of few percentiles of the number of new available testing kits per day. 

In Figure \ref{Fig: bistability delta}, we can see a similar bifurcation behavior between a full blow epidemics, and another regime in which epidemics diffusion stays limited and then vanishes, by fixing all parameters but $\delta$ which is freely moved. Again we observe a dramatic change of the epidemic diffusion in a very narrow range of the parameter $\delta$. These observations strongly suggest that the coupling between kits and the fractions of undetected and detected individuals is crucial for the possible evolution of the epidemics. 

\begin{figure}[t!]
\includegraphics[width=.91\columnwidth]{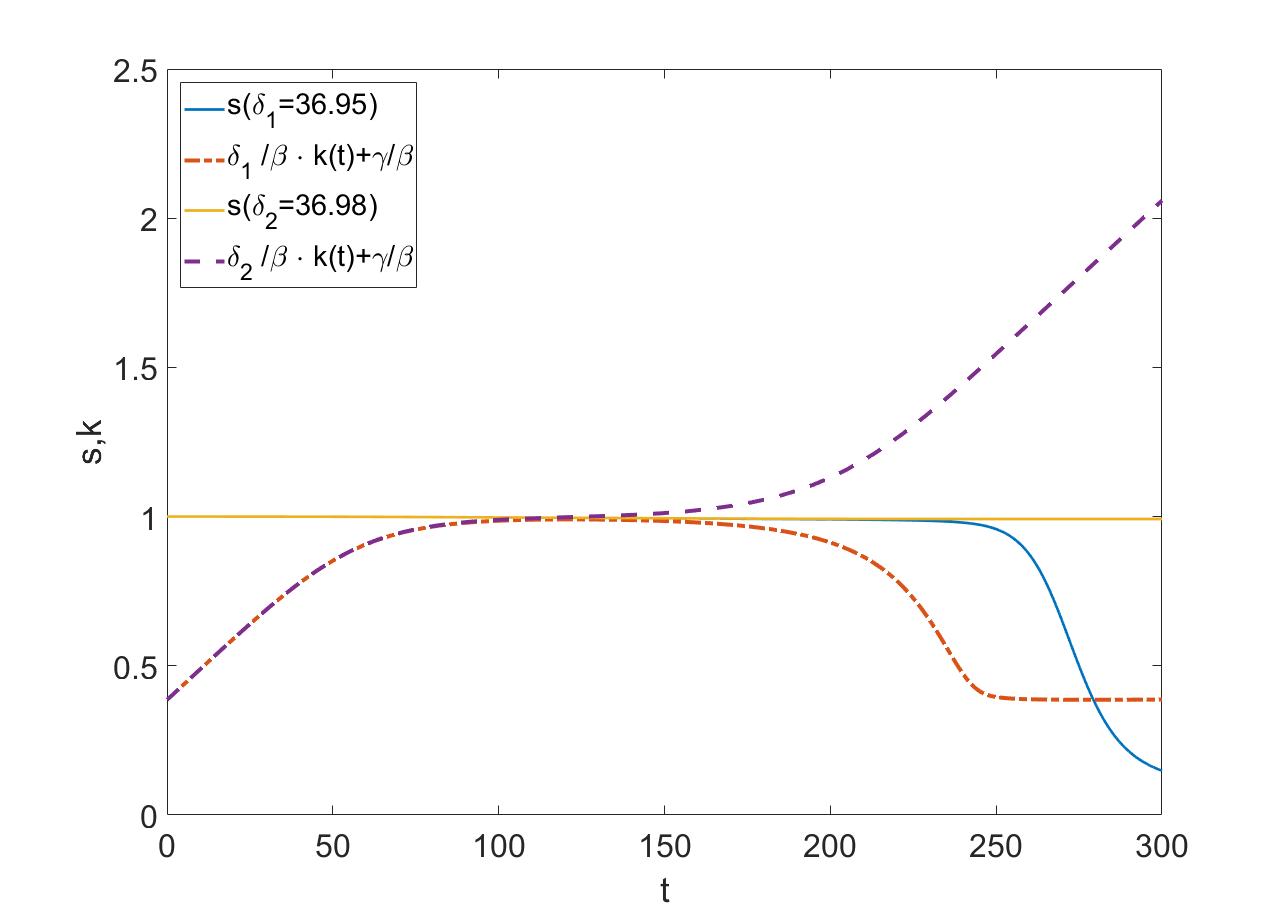}
\centering
\caption{We represent both $s(t)$ and $\frac{\delta}{\beta} k(t)+\frac{\gamma}{\beta}$ as function of time for the same parameters values of Figure \ref{Fig: bistability delta}: $\beta=\ln{(2)}/2.7$, $\gamma=0.099$, $\alpha=7.14\cdot10^{-5}$, $\epsilon=2$, and two choices of $\delta$, $\delta_1=36.95$ and $\delta_2=36.98$. The fraction $u(t)$ gets its maximum when the two curves cross. We see that for $\delta=\delta_2$ the infection is always well limited by the testing activity up to disappear, while for $\delta=\delta_1$ the epidemics explodes up to infect more half of population.}
\label{Fig: bistability explanation}
\end{figure}

Since a similar behavior is observed also for other kinds of coupling, introduced in Sect.~\ref{models}, we now give a more general argument on Eq.~\eqref{2} to explain this behavior. 
Let us start from the simplest case Eqs.~\eqref{1}-\eqref{5} and focus on the temporal location of the maximum of the fraction of undetected infected individuals. It is obtained by solving the equation $\dot{u}(t_c)=0$, which, through Eq.~\eqref{2}
\begin{align}
s(t_c) &= \frac{\delta}{\beta} k(t_c)+\frac{\gamma}{\beta}.
    \label{the equation}
\end{align}
In Fig. \ref{Fig: bistability explanation} 
we represent 
the right hand side of the Eq.~\eqref{the equation} as a function of time and $s(t)$ for the same two choices of the parameters of Figure \ref{Fig: bistability delta}, where all parameters coincide but $\delta$ for which we have two very close values $\delta_1=36.95$ and $\delta_2=36.98$ around the bifurcation point. We will have a maximum in $u(t)$ as soon as the curve of $\frac{\delta}{\beta} k(t)+\frac{\gamma}{\beta}$ crosses the curve of $s(t)$. We see that for $\delta=\delta_2$ the crossing time $100<t_c<150$ and happens for $s(t)\simeq 1$ so that the infection is strongly limited by the testing activity and $u(t)$ stays very small at all time up to vanish. On the contrary for $\delta=\delta_1$, $250<t_c<300$ and happens for $s(t)< 0.5$. This means that the infection exploded leading to a fraction larger than $0.5$ of infected people across the population. 
This is a further confirmation of the switching between the two aforementioned phases of the epidemics. 

In order to generally explain this transition, we make use of the general coupling term \eqref{FI} in the model equations. As long as the function $\phi$ is strictly positive and continuous we will have the same behavior, but with changed temporal location of the switch. In that case the switch will arise naturally by setting $\dot{u}(t_c)=0$ which, from Eq.~\eqref{2}, means through the solution $t_c$ of the equation :
\begin{align}
     \phi(u(t_c),k(t_c),\alpha,\delta) &= \beta s(t_c)  -\gamma
\end{align}

In order to proceed to a classification of the two phases, we have to study the second order time derivative of the fraction of undetected individuals $\ddot{u}$:
\begin{align}
    \ddot{u}(t_c)&=\beta \dot{s}(t_c) u(t_c)-u(t_c)\dot{\phi}(t_c)\,.\label{ddotu1}
\end{align}
Clearly $t_c$ will be the time of a local maximum if $\ddot{u}(t_c)<0$. This happens if
\begin{align}
    \beta\dot{s}(t_c)<\dot{\phi}(t_c)
    \label{ddotCond}
\end{align}
when the growth of the undetected sub-population is suppressed. Equation \eqref{ddotCond} says that the change of the rate $\beta s$ at which new undetected (i.e. new infected) are produced has to be smaller than the change of the rate $\phi$ at which new undetected are discovered by test and moved to the detected sub-population. 

\section{Discussion}
A simple  interpretation of this result is that when the rate 
of successful testing and the rate of recovery equals the rate of transmission of the infection (i.e. transformation of individuals from susceptible to the infected state), and the changes of this rates also coincide, the pandemic enters into a dynamical stationary state. Note that this does not mean that there are no newly infected, but simply that the number of new undetected per day is kept below a certain value. When the two rates equate, we have a clear separation between the region with small and manageable population of $u$ and a full blow up of the epidemics.
The repercussion of this result is that testing can have an immense impact if it is done in time, in a smart calibrated pace on the rate of transmission of the infection in the population, and tests are made available at a sufficient rate. Indeed it is important to stress that the way Singapore handled the Covid-19 crisis \cite{10.1001/jama.2020.2467}, at least in the first round of the infection, is very similar to our model. Moreover Japan and Hong Kong are also managing well the diffusion of the epidemics during the writing of this paper: indeed $\alpha=0.0002$, as reported for the Hong Kong case \cite{legido2020high}, is within the meaningful range of parameters we used in this model. This leads us to believe that those developed countries which are adopting testing policies postponing a widespread testing activity until they have full blown epidemics, put themselves in a very risky situation in which epidemics diffuse with poor control across the population. This result would also suggest that sharing of tests among nations is fundamental in order to mitigate the epidemics diffusion.

In the end we would like to once again stress that here we present toy model which is not calibrated and suitable to any kind of quantitative predictions. We believe that the testing strategy, and the modeling of detection of cases is of fundamental importance for the epidemics of COVID19 as well as for all possible future epidemics of unknown pathogens, and we hope this work can open the way to collaborate with institutions and researchers which are working on real testing to model it as best as possible.  


\section{Acknowledgements}
        
We want to express gratitude to Marko Popovi\'c, Bruno Marcos, Goran Duplan\v{c}i\'c, Zoltan Toroczkai, Fabio Franchini, Salvatore Marco Giampaolo and Antonio Scala for their useful comments. VZ and IB acknowledges partial support form QuantiXLie Centre of Excellence, a project co-fina nced by the Croatian Government and European Union through the European Regional Development Fund - the Competitiveness and Cohesion Operational Programme (Grant KK.01.1.1.01.0004, element leader N.P.). HS and VZ had their research supported by the European Regional Development Fund under the grant KK.01.1.1.01.0009 (DATACROSS). VZ also acknowledges the Croatian Science Foundation (HrZZ) Projects No. IP–2016–6–3347 and IP–2019–4–3321.

\bibliography{Biblio}   
%
%

\end{document}